\documentclass[12pt,twoside]{article}   
\usepackage[super,sort,comma]{natbib}

\usepackage{fancyhdr}		

\newcommand{\captionv}[3]{\begin{center}\parbox{#1cm}{\caption[#2]{{\sf #3}}}
        \end{center}}

\usepackage[section]{placeins}   %
\usepackage{graphicx}

\makeatletter \renewcommand\@biblabel[1]{$^{#1}$} \makeatother
\newcommand{\cen}[1]{\begin{center} #1 \end{center}}

\usepackage[mathlines]{lineno}

\usepackage{hyperref}
\hypersetup{ colorlinks,
    citecolor=blue,
    filecolor=blue,
    linkcolor=blue,
    urlcolor=blue
}

\usepackage{xcolor}

\definecolor{gray}{rgb}{0.6,0.6,0.6}
\definecolor{red}{rgb}{0.85,0,0}
\definecolor{green}{rgb}{0,0.85,0}
\definecolor{blue}{rgb}{0,0,0.85}
\definecolor{beige}{rgb}{0.92,0.87,0.78}
\usepackage[all]{hypcap}    

\begin{document}

\cen{\sf {\Large {\bfseries MRI Tissue Magnetism Quantification through Total Field Inversion with Deep Neural Networks } \\  
\vspace*{10mm}
Juan Liu, Kevin M. Koch} \\
Biomedical Engineering, Medical College of Wisconsin and Marquette University, Milwaukee, Wisconsin, 53226, USA\\
Radiology, Medical College of Wisconsin, Milwaukee, Wisconsin, 53226, USA
\vspace{5mm}\\
Version typeset \today\\
}

\pagenumbering{roman}
\setcounter{page}{1}
\pagestyle{plain}
email: kmkoch@mcw.edu \\

\begin{abstract}

\noindent {\bf Purpose:} Quantitative susceptibility mapping (QSM) utilizes MRI signal phase to infer estimates of local tissue magnetism (magnetic susceptibility), which has been shown useful to provide novel image contrast and as biomarkers of abnormal tissue. QSM requires addressing a challenging post-processing problem: filtering of image phase estimates and inversion of the phase to susceptibility relationship. A wide variety of quantification errors, robustness limitations, and artifacts plague QSM algorithms. To overcome these limitations, a robust deep-learning-based single-step QSM reconstruction approach is proposed and demonstrated. \\

{\bf Methods:} A deep convolutional neural network was trained to perform total field inversion (TFI) from input MRI phase images. This neural network was trained using magnetostatic physics simulations based on in-vivo data sources. Random perturbations were added to the physics simulations to provide sufficient input-label pairs for the training purposes. The network was quantitatively tested using gold-standard in-silico labeled datasets against established QSM total field inversion approaches. In addition, the algorithm was applied to susceptibility-weighted imaging (SWI) data collected on a cohort of clinical subjects with brain hemmhorage.\\

{\bf Results:} When quantitatively compared against gold-standard in-silico labels, the proposed algorithm outperformed the existing comparable approaches.   High quality QSM were consistently estimated from clinical susceptibility-weighted data on 100 subjects without any noticeable inversion failures.\\

{\bf Conclusions:} The proposed approach was able to robustly generate high quality QSM with improved accuracy in in-silico gold-standard experiments. QSM produced by the proposed method can be generated in real-time on existing MRI scanner platforms and provide enhanced visualization and quantification of magnetism-based tissue contrasts.

\end{abstract}

\newpage     


\newpage     
\setlength{\baselineskip}{0.7cm}      

\pagenumbering{arabic}
\setcounter{page}{1}
\pagestyle{fancy}

\section{Introduction}
Magnetic susceptibility, which is a parametric measure of material magnetism~\cite{schenck1996role}, has long been utilized to generate diagnostic imaging contrast through magnetic resonance imaging (MRI) techniques such as T$_2^*$ or enhanced susceptibility-weighted imaging (SWI)~\cite{haacke2004susceptibility}.
Quantitative Susceptibility Mapping (QSM) is a MRI post-processing technique that expands upon these widely utilized susceptibility-weighted practices to provide estimates of underlying tissue magnetism~\cite{wang2015quantitative}.

SWI is widely-used for clinical assessments, including traumatic brain injury\cite{babikian2005susceptibility,huang2015susceptibility}, vascular malformations~\cite{thomas2008clinical,de2008susceptibility}, brain tumors~\cite{sehgal2006susceptibility, rauscher2005magnetic}, neurodegenerative diseases~\cite{haacke2009characterizing,sehgal2005clinical}, etc. Despite its widespread utility and usage, SWI suffers from blooming intensity artifacts and cannot differentiate calcifications from small microbleeds or iron deposits. Besides, SWI contrasts change with field strength, echo time settings, and depend on both the shape and the orientation of the blood vessels. By comparison, QSM quantitatively estimating the underlying tissue magnetic susceptibility is able to overcome the limitations of SWI. Existing studies have explored to usage of QSM to quantification of specific biomarkers including iron, calcium, gadolinium etc \cite{haacke2005imaging,deistung2013toward,zhang2015quantitative,zheng2013measuring}. QSM has been explored to study brain tumors\cite{deistung2013quantitative,wisnieff2015quantitative}, neurodegenerative disorders and multiple sclerosis\cite{bilgic2012mri, lotfipour2012high,wisnieff2015quantitative}, iron overload in the liver\cite{sharma2015quantitative}, blood oxygenation assessment\cite{fan2014quantitative}, and mild traumatic brain injury\cite{koch2018quantitative} as well.

QSM is performed by collecting phase-sensitive MRI, estimating local magnetic perturbations along the direction of the polarizing \textbf{B}$_0$ magnetic field, and then using magnetic perturbations to estimate source magnetic susceptibility (i.e. magnetism) within tissue~\cite{wang2015quantitative}.  
The collection of phase-sensitive MRI and associated steps to develop local field perturbation estimates is well-established.   On the contrary, the transformation of raw local field measurements to tissue susceptibility estimates is fraught with technical challenges.  Historically, QSM transformation from raw perturbation fields usually consist of two steps: (i) background field removal to determine the local tissue field, and (ii) inversion from the local field to the tissue susceptibility. Both steps require solving ill-posed inverse computational problems. 

Background field removal (BFR) methods typically rely on spatial filtering and/or dipole field modelling methods~\cite{schweser2011quantitative, sun2014background, liu2011novel, zhou2014background}. Existing BFR approaches can often perform well under ideal circumstances with perfect brain tissue masking.  However, in the presence of these confounds, BFR is often compromised with residual background field and image artifacts in the local field, which then leads to artifacts and inaccuracies in final QSM estimates. 



To avoid BFR error propagation into QSM estimation, several single-step QSM algorithms, which directly estimate the susceptibility distribution from the raw perturbation field, have been proposed.  These approaches,  hereby denoted as ``total field inversion" (TFI) methods, utilize algorithmic adjustments such as Laplacian spatial conditioning operators~\cite{Berkin2014SingleStep, Kristian2014SingleStep, Tian2014Differential, Samir2014SingleStep}, preconditioning boosted iterative inversions~\cite{liu2017preconditioned}, or total variation (TV) and/or Tikhonov regularized inversion~\cite{chatnuntawech2017single,sun2018whole} to address the challenges imposed by TFI.   

TFI methods have shown improved susceptibility estimation performance compared to conventional two-step QSM methods.   However, all existing TFI methods require tuning of regularization parameters, which introduce estimation biases and artifacts. In addition, these methods require substantial computation times to solve their respected iterative optimization problems. Finally, our investigations have found that these methods still exhibit residual streaking artifacts and BFR failures in regions with large susceptibility variations, such as intracranial hemorrhages. Furthermore, we have found existing available methods to have compromised performance when applied to routinely-acquired clinical susceptibility-weighted imaging data, which is highly non-isotropic in resolution (due to examination time constraints).  

To address the limitations of existing TFI methods, here we demonstrate a deep-learning-based TFI approach. Leveraging the well-defined forward susceptibility-to-field physics model~\cite{marques2005application,salomir2003fast}, synthetic data are used for neural network training. This addresses the common machine-learning challenge of collecting and curating accurate ground-truth training data. The proposed TFI deep learning network is demonstrated on source MRI data collected under two different use cases with substantially different image resolutions.  Quantitative in-silico gold-standard performance evaluations of the proposed method and several existing TFI and multi-step QSM algorithms are also presented.  In addition, several qualitative demonstrations are presented to show superior performance of the proposed method relative to existing approaches.    

\section{Methods}

\subsection{Training Data} 

Multiple-Orientation QSM datasets, such as Calculation of Susceptibility through Multiple Orientation Sampling (COSMOS)~\cite{liu2009calculation} or Susceptibility Tensor Imaging (STI)~\cite{liu2010susceptibility}, which are not clinically viable acquisition approaches, are often treated as QSM golden-standard estimates.   While these approaches do address concerns of artifacts and quantification concerns with the field inversion step, there are several concerns to consider when using such data for QSM deep learning training labels.  First, the use of these approaches is assuming perfect background field removal of the training data.  Second, these techniques are iterative regularized algorithms that are not guaranteed to provide exact estimates of the field-to-source magnetism relationship (i.e the ``input" to ``label" relationship is not a true ``gold standard").   Rather than relying on such acquisitions for neural network training, we propose that the well-defined forward source-to-field model offers a more flexible and accurate gold-standard training label.  In our approach, we utilize cohorts of \emph{in-vivo} QSM estimates as an input to a computational training data generator, which allows for trivial data augmentation practices to provide large numbers of realistic training inputs with true gold-standard labels accurate to levels of machine precision.  In this approach, the accuracy of the original \emph{in-vivo} QSM estimate is largely irrelevant, as the network is being trained exclusively by the ensuing closed-form forward source-to-field computation.     

200 \emph{in vivo} QSM datasets were used as inputs to the simulated neural network training data generator. The resolution of this source data was 0.5x0.5x2.0 mm$^3$. From the four echo time images in each data set, QSM estimates were generated using the following existing tools: brain masking using SPM~\cite{ashburner2005unified}, BFR using the Regularization-enabled Sophisticated Harmonic Artifact Reduction on Phase data (RESHARP) \cite{sun2014background} method, and susceptibility inversion was performed using a previously developed Approximated Susceptibility through Parcellated Encoder-decoder Networks (ASPEN)~\cite{liu2019quantitative}.

\begin{figure}[ht]
\begin{center}
\includegraphics[width=14cm]{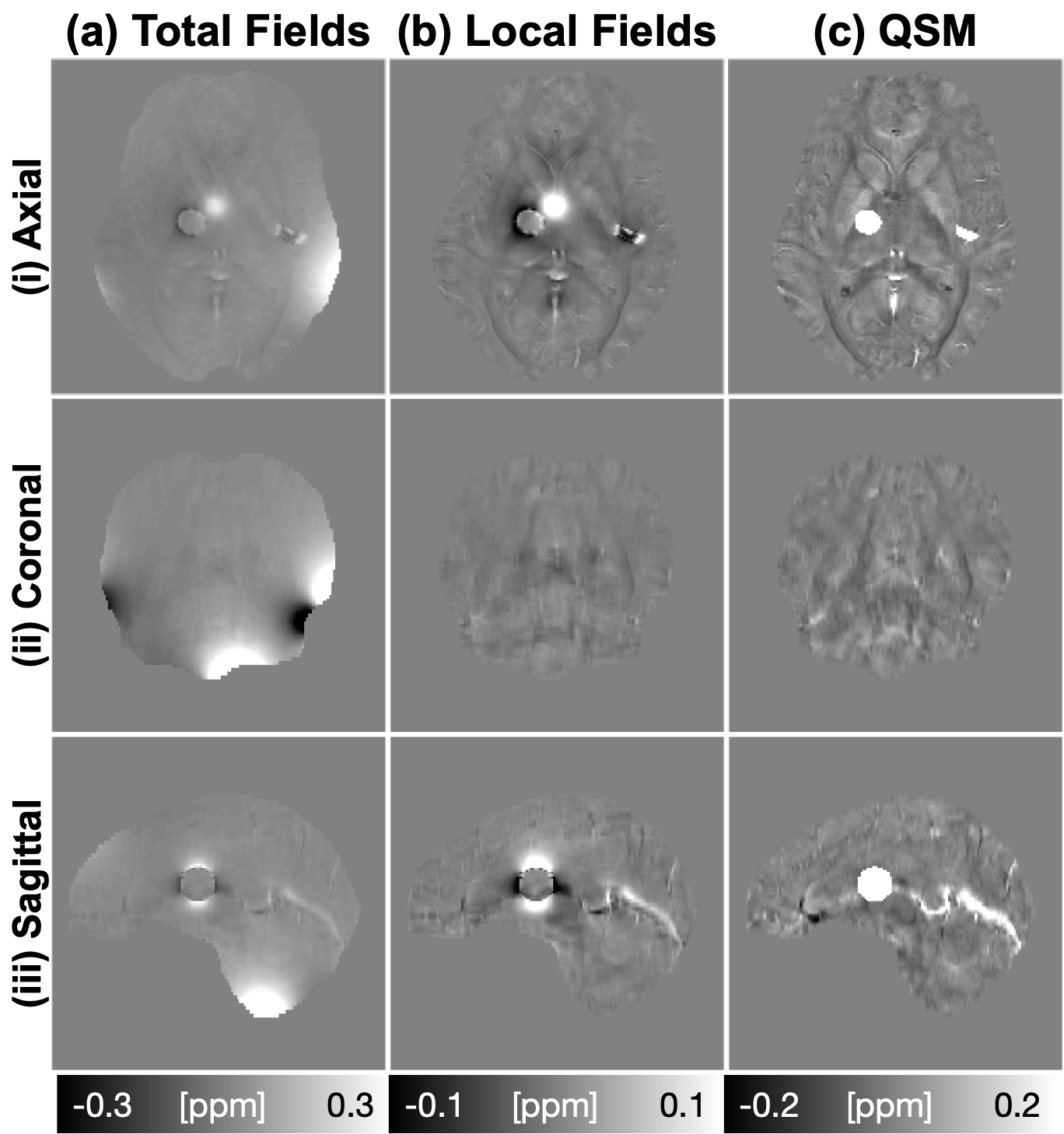}
\captionv{12}{}{Illustration of training datasets.
\label{fig_trainingdata} 
}
\end{center}
\end{figure}

Besides, geometric shapes such as ellipsoid, sphere, cuboid and cylinder with random susceptibility values and random orientations are randomly placed on the susceptibility maps. Using the \emph{in-vivo} QSM estimates, local tissue magnetic perturbations were calculated using well-known dipole convolution methods~\cite{salomir2003fast,marques2005application}. The background perturbation fields were then simulated from random magnetic susceptibility sources to mimic background field. The field perturbation for training total field inversion was then constructed from the superimposed local tissue perturbations and the background perturbation. Fig.~\ref{fig_trainingdata} provides an example input (total field perturbations) and label (tissue susceptibility) from this training data construction process.

\subsection{Neural Network Architecture and Training} 

A 3D convolutional neural network with encoder-decoder architecture was trained to perform total field QSM inversion with 3D unwrapped total field maps and brain masks as the inputs. In vanilla convolutional neural networks, the spatial invariant kernels is inappropriate to solve our problem due to large background fields close to brain boundary and invalid data outside brain mask. To overcome these limitations, gated convolution with LeakyReLU as activation function and Sigmoid for gating value was applied for automatically selecting adaptive features for each channel and each spatial location. Dilated gated convolution was applied in deeper convolutional layers to increase the receptive fields. Moreover, a non-local block was inserted in encoder-decoder network to capture long-range dependencies for non-local susceptibility estimation\cite{wang2018non}.

For performance evaluation using synthetic data and QSM challenge, the image resolution is 1.06 mm$^3$ isotropic with neural network input size 160x160x160. For clinical data, the image resolution is 0.76x0.76x3.0 mm$^3$ with size 256x256x64. 5000 data were used for training. L1 loss between the label and output was utilized as a loss function. The RMSprop optimizer was used in the deep learning training. The initial learning rate was set as 0.0001, with exponential decay at every 200 steps. Two NVIDIA tesla k40 graphics processing units (GPUs) were used for training with batch size 2. The neural network was trained and evaluated using Keras with Tensorflow as backend.

\subsection{Performance Evaluation}

\subsubsection{Synthetic Data} 

100 simulated data sets generated in similar fashion to the training data without containing randomly inserted geometric shapes were used to evaluate the performance of QSMTFINet compared with both single-step and two-steps QSM reconstruction approaches, were selected for comparison, including (i) single step total variation QSM (SS-TV-QSM) \cite{chatnuntawech2017single}, (ii) least-norm QSM (LN-QSM) \cite{sun2018whole}, (iii) RESHARP and Thresholded K-space Division (TKD), (iv) Projection onto Dipole Fields (PDF) \cite{liu2011novel} and Morphology Enabled Dipole Inversion (MEDI) \cite{liu2012morphology}.

Estimation errors from each technique were computed using root mean squared error (RMSE), high-frequency error norm (HFEN), and structural similarity (SSIM) index with the ground truth susceptibility maps. 

\subsubsection{2016 QSM Reconstruction Challenge Dataset} 

QSM reconstruction challenge dataset was acquired from a healthy 30-year-old female subject. Data for COSMOS and susceptibility-tensor computation was computed using a heavily accelerated wave-CAIPI~\cite{bilgic2015wave} acquisition at 1.06 mm isotropic resolution collected at 12 different head orientations at a single echo time~\cite{langkammer2018quantitative}. 

Using this gold-standard evaluation dataset, the proposed method was compared to SS-TV-QSM, LN-QSM, TKD, and MEDI approaches. TKD and MEDI were performed using the provided local tissue field. TKD results were provided publically by the QSM challenge organizers, with threshold 0.19 which yields the best trade-off between quantification accuracy and artifacts. All methods were evaluated against the ``gold standard" STI (3,3) component computational result provided with the challenge dataset~\cite{liu2010susceptibility}. 

\subsubsection{Clinical Data} 

One hundred clinical QSM data were acquired using gradient echo T$_2^*$ weighted angiography (SWAN, GE), a new method for SWI with short acquisition times, at a 3T MRI scanner (GE Healthcare MR750) with data acquisition parameters: in-plane data acquisition matrix 288x224, field of view 22 cm, slice thickness 3 mm, autocalibrated parallel imaging factors 2x1, number of slices 46-54, first echo time 12.6 ms, echo spacing 4.1 ms, number of echoes 7, flip angle 15$^o$, repetition time 39.7 ms, total scan time about 2 minutes.  

The SWI images were processed by vendor reconstruction algorithms. The raw k-space data were saved for offline QSM processing. Multi-echo real and imaginary data were reconstructed from k-space data, with reconstruction matrix size 288x288, voxel size 0.76x0.76x3.0 mm$^3$. The field map was obtained by the fitting of multi-echo phases. Brain masks were obtained using the FSL brain extraction tool.

Using the total field map and brain mask, representative state-of-the-arts methods, both two-steps QSM reconstruction and single-step QSM reconstruction, were selected for comparison, including the five QSM reconstruction methods mentioned above. Three clinical cases were chosen to demonstrate TFINet performance. 

For the purposes of performance evaluation, with RESHARP, spherical kernel radius was set as 6mm to trade off the background removal performance and brain erosion; for TKD, the threshold was set to 0.20; for MEDI, the regularization factor was set to 1000, which can yield a good trade-off between quantification accuracy and artifacts; the TV and Tikhonov regularization parameters for LN-QSM were set to 4x10$^{-4}$ and 10$^{-3}$ respectively. MEDI toolbox, SS-TV-QSM code, LN-QSM code publicly provided by the authors were used to calculate the QSM images\cite{MEDIcornellWEB,berkinMITWEB,LNQSMCode}.

\section{Results}

\subsubsection{Synthetic Data} 

\begin{figure}[ht]
\begin{center}
\includegraphics[width=14cm]{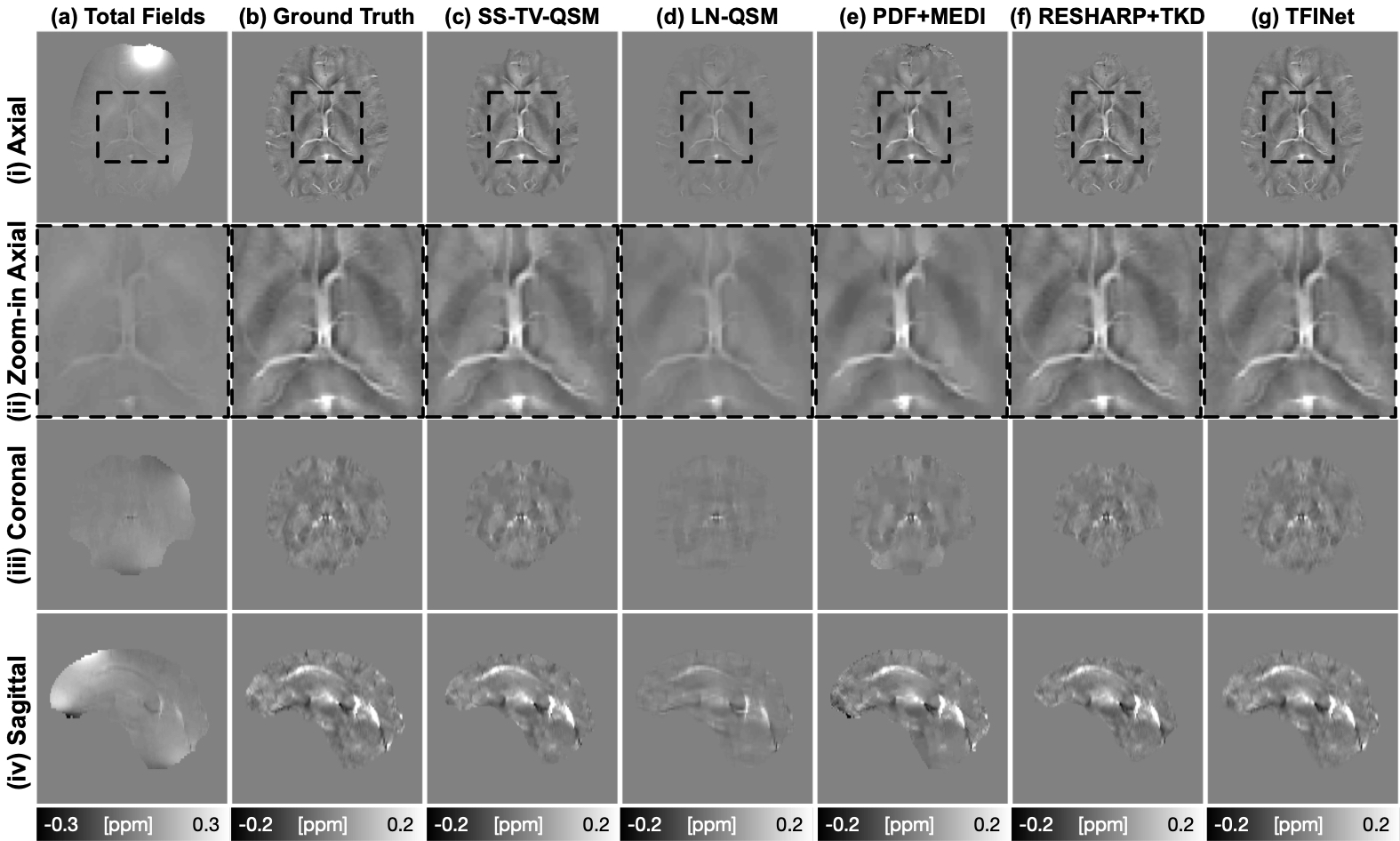}
\captionv{12}{}{Total field (a), ground truth QSM images (b) and estimated QSM images (c-g) by five methods from one of synthetic test data in axial (i), coronal (iii), and sagittal (iv) views.
\label{fig_testdata} }
\end{center}
\end{figure}

\begin{figure}[ht]
\begin{center}
\includegraphics[width=14cm]{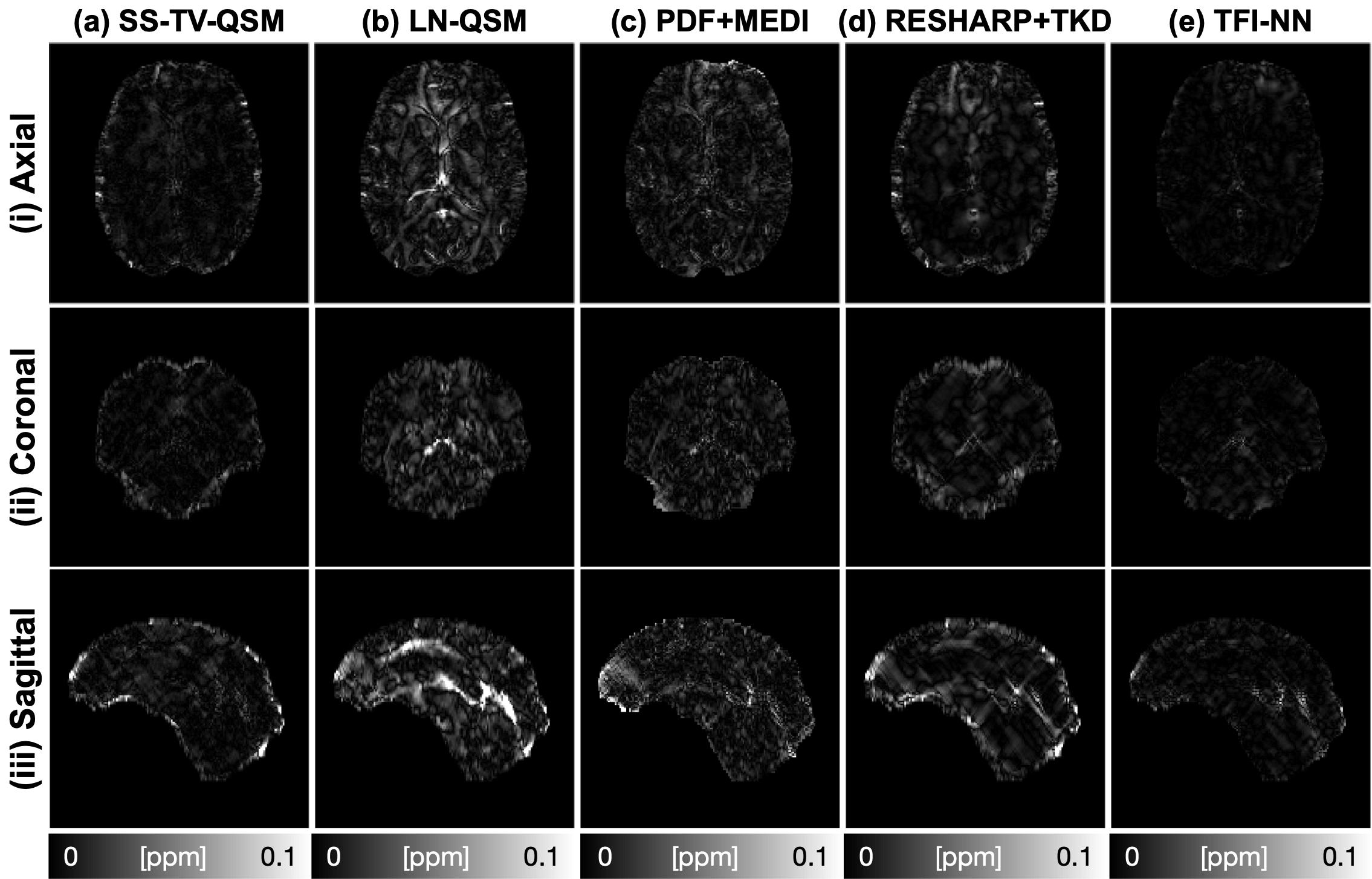}
\captionv{12}{}{The residual error maps of Fig.\ref{fig_testdata}. TFINet results have the least errors compared with all other results.
\label{fig_testdata_errmap} }
\end{center}
\end{figure}

Table~\ref{table_sim_data_normal} illustrates the RMSE, HFEN, and SSIM using five QSM reconstruction methods compared with ground truth from 100 simulated datasets. The proposed method achieved the best score in RMSE, HFEN, and SSIM. In Fig.\ref{fig_testdata}, QSM images reconstructed by five different reconstruction methods are compared with ground truth susceptibility maps. The residual maps compared with ground truth are shown in Fig.\ref{fig_testdata_errmap}. Based on visual assessment, TFICnn results have best susceptibility estimations and least residual errors.  

\subsubsection{2016 QSM Reconstruction Challenge Dataset} 

\begin{figure}[ht]
\begin{center}
\includegraphics[width=14cm]{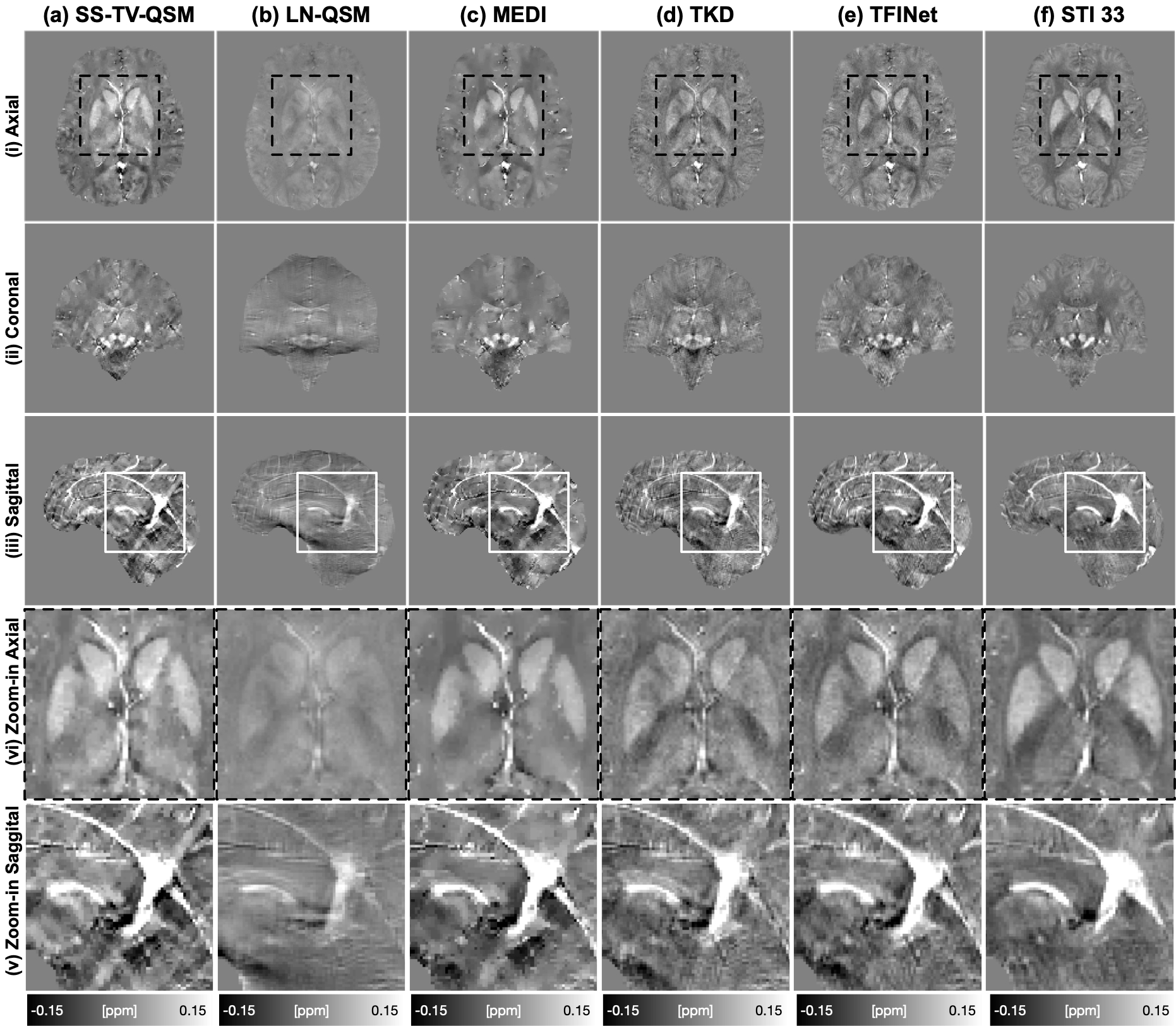}
\captionv{12}{}{STI (3,3), TKD, SS-TV-QSM, LN-QSM, and TFINet susceptibility maps from 2016 QSM reconstruction challenge in axial (i), zoom-in axial (ii), coronal (iii), and sagittal (iv) views.
\label{fig_qsmchallenge} }
\end{center}
\end{figure}

Fig.~\ref{fig_qsmchallenge} illustrates susceptibility maps from the 2016 ISMRM QSM Challenge dataset reconstructed using five methods, displayed in 3 reformatted planes(rows i, ii, iii). Streaking artifacts are clearly identified in the sagittal reformat for SS-TV-QSM maps. In addition, the SS-TV-QSM map shows clearly compromised spatial resolution relative to the other maps. These observations are amplified in the zoomed maps in the last two rows (iv, v), where clear performance improvements of the proposed method in reproducing the fine structure of the STI (3,3) map and no-visible streaking artifacts.

\subsubsection{Clinical Data}

\begin{figure}[ht]
\begin{center}
\includegraphics[width=14cm]{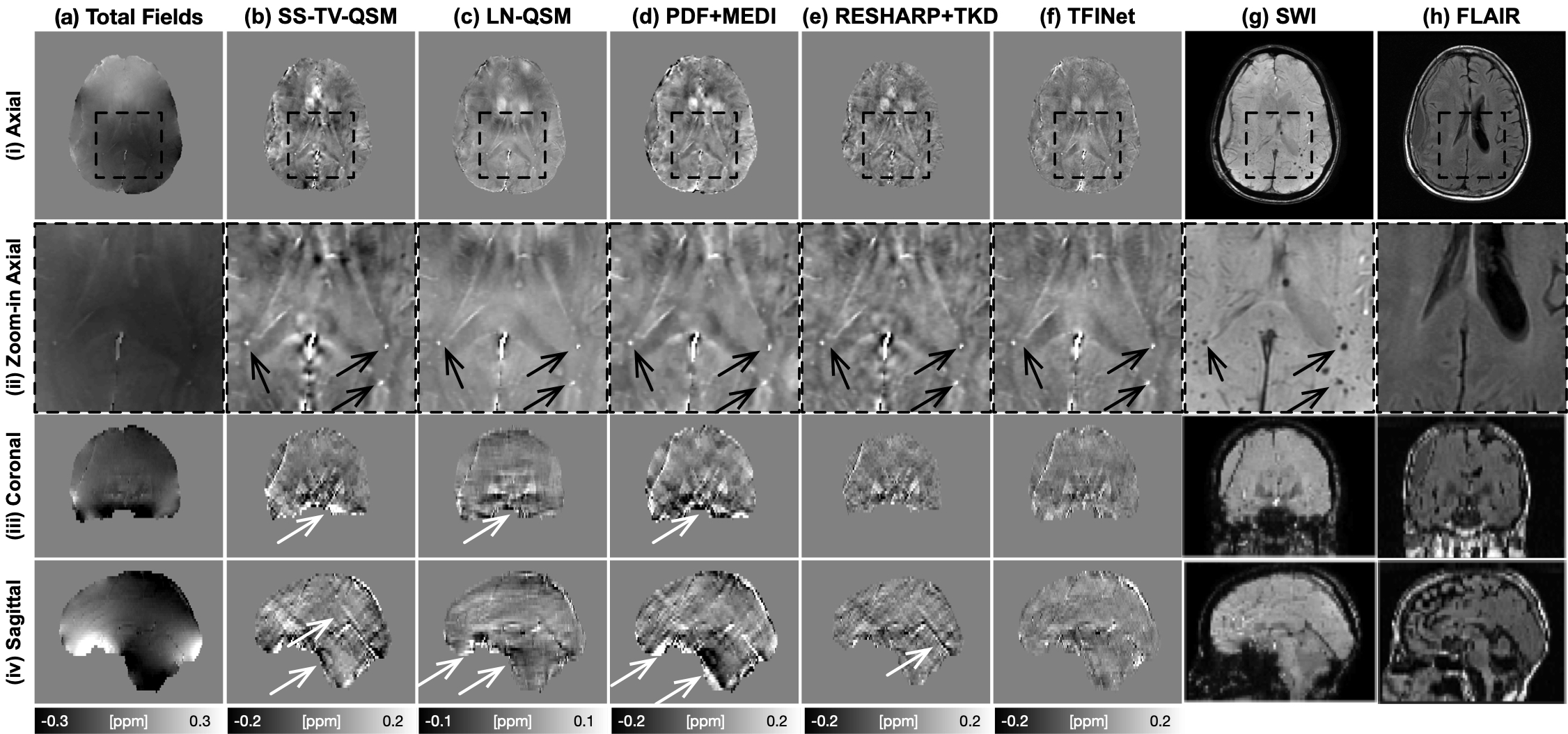}
\captionv{12}{}{Total fields (a), QSM images (b-f), SWI images (g), and FLAIR images (h) from a 65-year-old patient with stroke. In zoom-in axial (ii), a few hypointense regions of SWI image (black arrows) is hyperintense in QSM, indicating it iron deposition or hemorrhage. In saggital plane (iv), streaking and shading artifacts are clearly visible in SS-TV-QSM, LN-QSM, PDF+MEDI, and RESHARP+TKD images. TFINet can greatly suppress the streaking artifacts.
\label{fig_p597} }
\end{center}
\end{figure}

In Fig.~\ref{fig_p597}, QSM results on one 65-year-old patient with stroke is shown in three (axial/coronal/sagittal) views. In zoom-in axial (ii), a few hypointense regions of SWI image (black arrows) are hyperintense in QSM, indicating it iron deposition or hemorrhage. In SS-TV-QSM results, obvious image blurring is shown in axial plane and apparent streaking artifacts in coronal and saggital planes, as shown (c, i-iv). Though LN-QSM results show good image sharpness and non-visible streaking artifacts, it suffers image intensity variation across the brain volume and shading artifacts near the brain boundary, as shown (d, i-iv). The QSM images of PDF+MEDI show good image quality, yet with visible image blurring and streaking artifacts, as shown (e, i-iv). RESHARP+TKD results suffer from brain erosion, image blurring, and streaking artifacts, as shown (f, i-iv). Compared with other methods, TFINet results (f, i-iv) show fine details and negligible streaking artifacts. 


In Fig.~\ref{fig_p1619}, QSM and SWI images of a 56-year-old subject with hemorrhagic intracranial metastases is illustrated. SS-TV-QSM, PDF+MEDI, and RESHARP+TKD suffers severe shading artifacts and streaking artifacts, as shown (b, d, e, i-iv). LN-QSM shows strong shading artifacts, as shown (c, i-iv). With comparison to other methods, TFINet results (f, i-iv) showed improved image sharpness, clear tissue structures, and no shading and streaking artifacts around bleeding regions. Besides, two small calcification is dark/hypointense on SWI images and diamagnetic on TFINet QSM images (white dash arrows).

\begin{figure}[ht]
\begin{center}
\includegraphics[width=14cm]{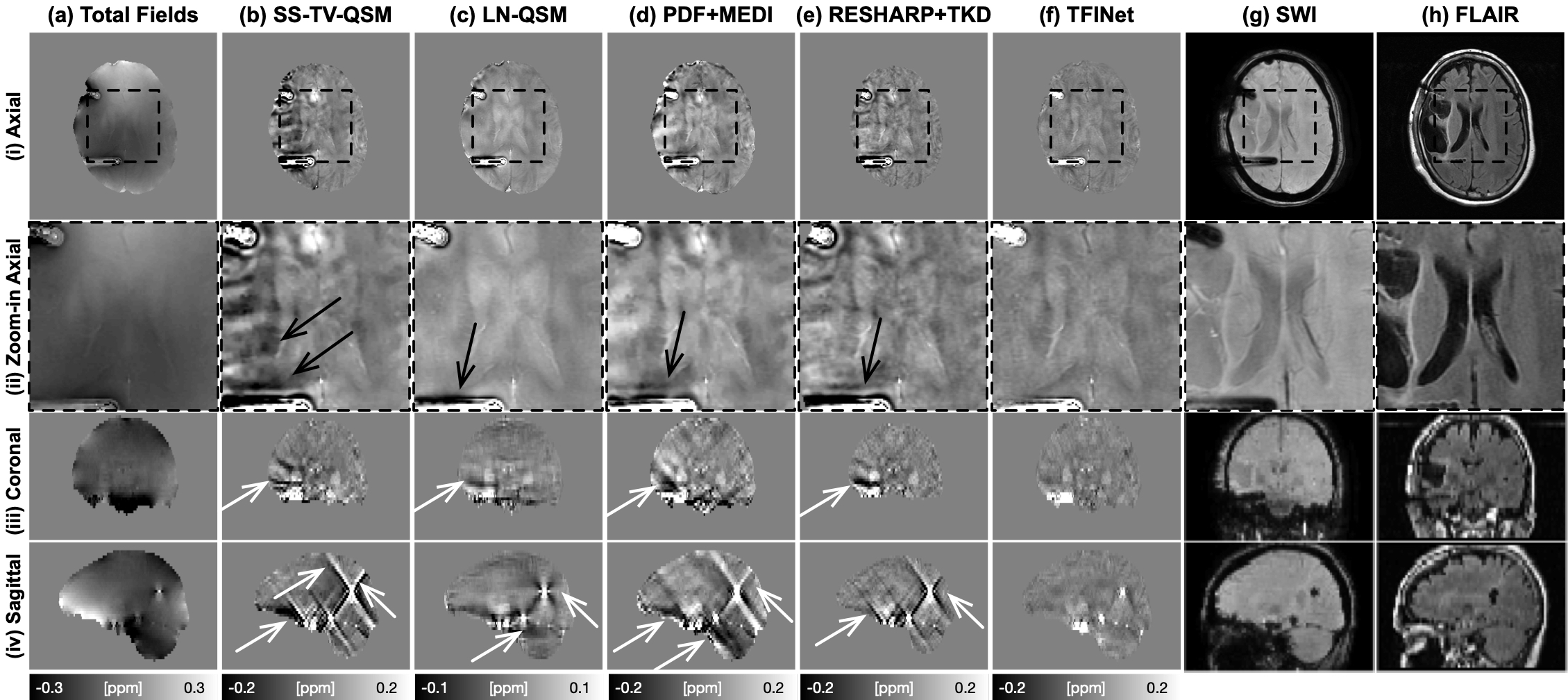}
\captionv{12}{}{Total fields (a), QSM images (b-f), SWI images (g), and FLAIR images (h) from a 37-year-old patient with surgical planning. In zoom-in axial (ii), image blurring and shading artifacts is clearly visible in SS-TV-QSM, PDF+MEDI, and RESHARP+TKD images. In coronal and saggital plane (iii, iv), streaking and shading artifacts are clearly visible in SS-TV-QSM, LN-QSM, PDF+MEDI, and RESHARP+TKD images. TFINet results show well-preserved microstrctures and non-visible artifacts.
\label{fig_p548}}
\end{center}
\end{figure}

In Fig.~\ref{fig_p548}, QSM results and SWI images are shown from a 37-year-old subject with surgical planning in three (axial/coronal/sagittal) views. Close to the bleeding regions, the SS-TV-QSM images show severe image blurring and shading artifacts in axial plane and severe streaking artifacts in coronal/saggital planes, as shown (b, i-iv). LN-QSM results clearly show black shading artifacts close to the intracerebral hemorrhage region and brain boundaries, as shown (c, i-iv). PDF+MEDI and RESHARP+TKD results show blurring in axial view, and apparent streaking artifacts in coronal/sagittal views, as shown (d, e, i-iv). Compared with all other methods, TFINet can produce high quality QSM images with fine details and non-visible streaking artifacts, shown in (f, i-iv). 

\begin{figure}[ht]
\begin{center}
\includegraphics[width=14cm]{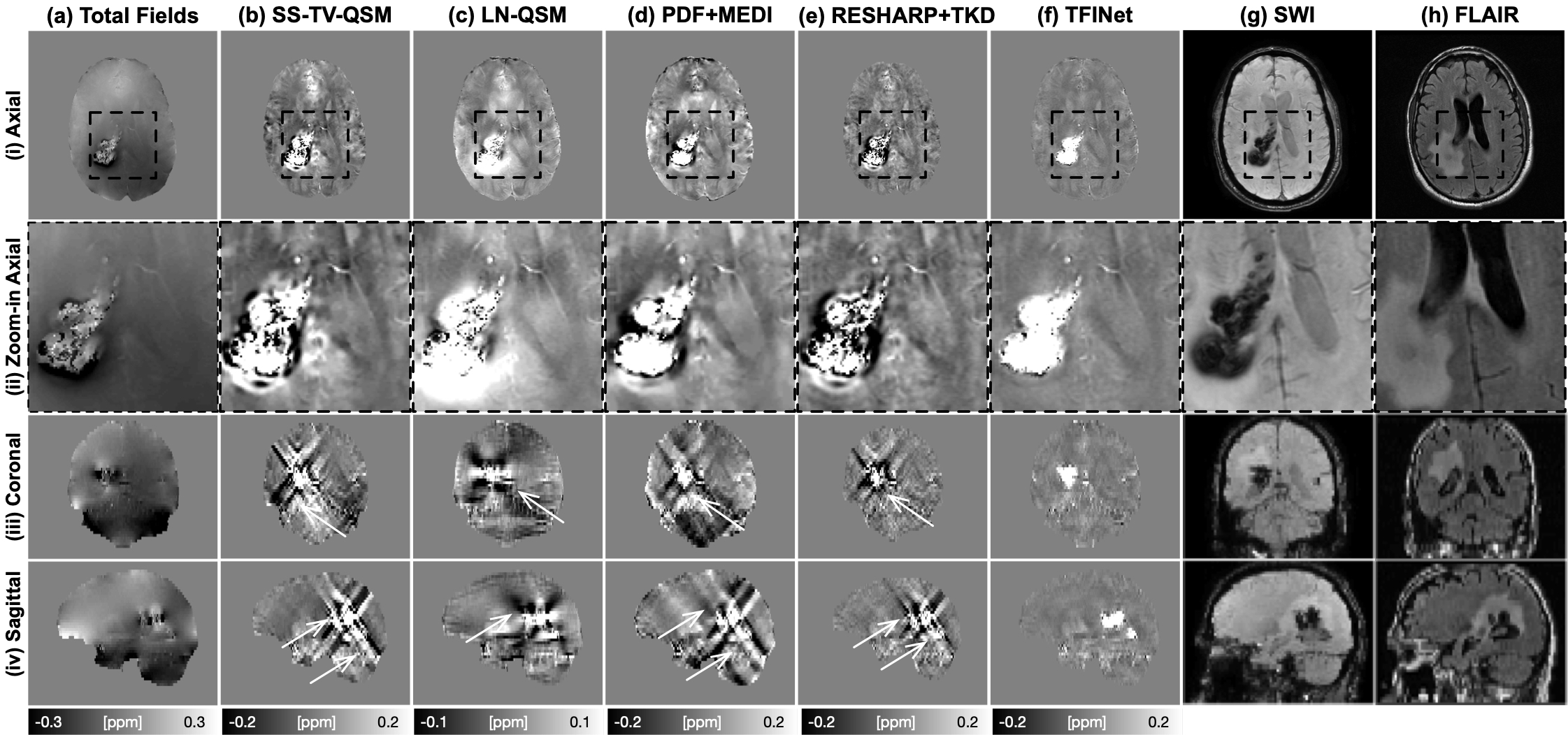}
\captionv{12}{}{Total fields (a), QSM images (b-f), SWI images (g), and FLAIR images (h) from from a  54-year-old  subject  with  poststereotactic radiosurgery (SRS) brain metastasis. In zoom-in axial (ii), image blurring and shading artifacts is clearly visible in SS-TV-QSM, PDF+MEDI, and RESHARP+TKD images. In coronal and saggital plane (iii, iv), streaking and shading artifacts are clearly visible in SS-TV-QSM, LN-QSM, PDF+MEDI, and RESHARP+TKD images. TFINet results show well-preserved microstrctures and non-visible artifacts.
\label{fig_p627}}
\end{center}
\end{figure}

In Fig.~\ref{fig_p627}, QSM results and SWI images are shown from from a  54-year-old  subject  with  poststereotactic radiosurgery (SRS) brain metastasis. Close to the bleeding regions, SS-TV-QSM, LN-QSM, PDF+MEDI, and RESHARP+TKD results (b-e) shows severe shading and bleeding artifacts. Compared with all other methods, TFINet can produce high quality QSM images with fine details and non-visible streaking artifacts, shown in (f, i-iv). 

\section{Discussion} 

In this work, TFINet was evaluated on synthetic data, QSM challenge dataset, and clinical dataset. Quantitative metrics and visual assessment demonstrated that TFINet outperform existing single-step QSM inversion and two-steps QSM inversion approaches. TFINet results have improved image sharpness, well-preserved microstructures, and non-visible streaking artifacts.

With comparison with single-step QSM methods SS-TV-QSM and LN-QSM, TFINet can produce high quality QSM images with better image sharpness and invisible image artifacts. SS-TV-QSM and LN-QSM use regularization methods to solve the ill-posed inverse problem, which require careful regularization parameter tuning and long iterative computation times. Though they eliminate the background field removal step, the QSM images suffer from over-smoothing, streaking artifacts, shading artifacts, and large susceptibility quantification errors close to intracerebral hemorrhage regions. The results show that TFINet can overcome the limitations and greatly improve susceptibility quantification accuracy. 

Compared with two-steps QSM methods, TFINet can not only speed up the QSM processing but also eliminate the background field removal error propagation into susceptibility estimation. Besides, the proposed methods overcome the constraints of existing background removal approaches, such as brain erosion, parameter tuning and image artifacts. In Fig.~\ref{fig_demo}, the PDF+MEDI suffers from image blurring and streaking artifacts, especially large susceptibility errors and streaking artifacts near the brain boundary and brain hemorrhage regions, which is partly due to inaccurate background removal close to brain boundary in PDF.   

The presented TFINet approach introduces several important innovations. First, it performs whole brain high-resolution QSM inversion using a neural network, which can avoid patch merging and tiling artifacts. Second, it utilizes a non-local block to increase the receptive fields and capture long-range information for non-local susceptibility estimation. Third, it uses gated convolutions to learn spatial information to spatially adaptive perform inner brain and close to brain boundary susceptibility estimation. 

\begin{figure}[ht]
\begin{center}
\includegraphics[width=14cm]{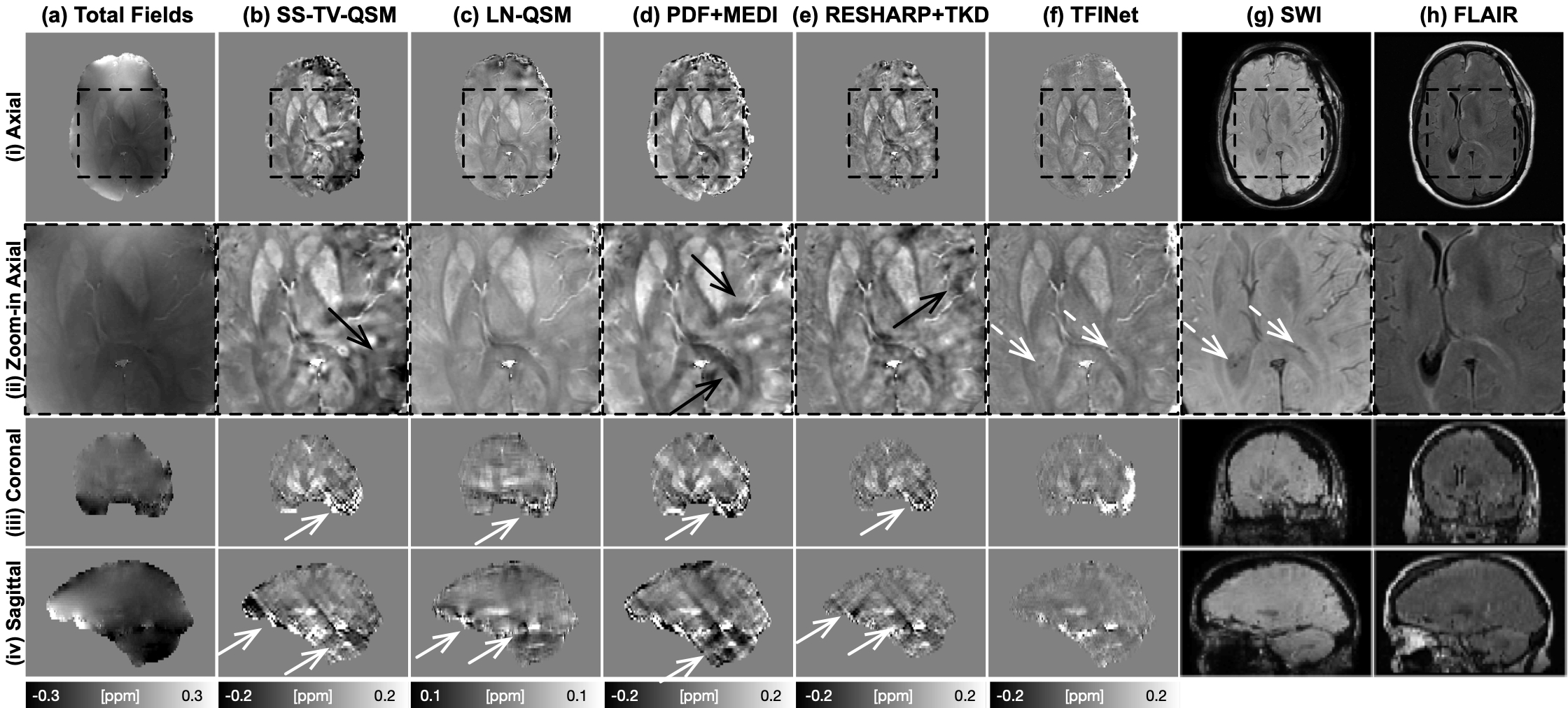}
\captionv{12}{}{Total fields (a), QSM images (b-f), SWI images (g), and FLAIR images (h) from a 56-year-old subject with hemorrhagic intracranial metastases. In zoom-in axial (ii), image blurring and shading artifacts is clearly visible in SS-TV-QSM, PDF+MEDI, and RESHARP+TKD images. In TFINet, two small calcification (white dash arrows) is hypointense on SWI image and diamagnetic on QSM image. In coronal and saggital plane (iii, iv), streaking and shading artifacts are clearly visible in SS-TV-QSM, LN-QSM, PDF+MEDI, and RESHARP+TKD images. TFINet can greatly suppress the streaking artifacts.
\label{fig_p1619}
}
\end{center}
\end{figure}

This feasibility study has also demonstrated the ability to use existing standard of SWI raw data to reconstruct QSM for clinical utility. This offers the possibility of QSM use in clinical operation without any additional scans beyond current standard of care protocols. Combining SWI magnitude and QSM estimation images may offer new diagnostic capabilities to assist radiological interpretation. In particular, it is well-known that SWI suffers from blooming artifacts and difficulties in differentiating  calcifications and hemosiderin. QSM can overcome these limitations of SWI, which can expand the roles of SWI and QSM in neuroradiology clinical and research arenas. In Fig.\ref{fig_p1619}, the calcification is easily differentiated in QSM maps. From Fig.\ref{fig_p1619} and Fig.\ref{fig_p548}, TFINet results show no shading artifacts or streaking artifacts around the lesions, while also  preserving the details of fine structures.

\section{Conclusion} 
In summary, a deep-learning-based single-step QSM approach have been demonstrated. It can substantially improve brain susceptibility estimation using clinical QSM data. This capability opens up a wide array of QSM investigations using clinically acquired SWI data to derive QSM maps across a host of neuroimaging indications.

\newpage     
	     
\section{Tables}

\begin{table}[!ht]
\centering
\captionv{12}{}{Numerical measures of QSM reconstruction quality on 100 synthetic data.
\label{table_sim_data_normal}
\vspace*{2ex}}
\begin{tabular}{|c|c|c|c|c|c|}
\hline
\multicolumn{1}{|c}{} & 
\multicolumn{1}{|c}{SS-TV-QSM} & 
\multicolumn{1}{|c}{LN-QSM} &
\multicolumn{1}{|c}{PDF+MEDI} &
\multicolumn{1}{|c}{RESHARP+TKD} & 
\multicolumn{1}{|c|}{TFINet}\\
\hline 
{{RMSE} ($\%$)} & 
{$46.4\pm1.9$} & 
{$78.8\pm7.4$} & 
{$67.7\pm6.5$} & 
{$59.7\pm2.3$} & 
{\bf{$28.5\pm3.1$}} \\
\hline
{{HFEN} ($\%$)} & 
{$42.6\pm2.2$} & 
{$78.2\pm7.7$} & 
{$54.1\pm5.6$} & 
{$58.8\pm2.6$} & 
{\bf{$26.5\pm2.4$}} \\
\hline 
\multicolumn{1}{|c}{{SSIM} (0-1)} & 
\multicolumn{1}{|c}{$0.832\pm0.008$} & 
\multicolumn{1}{|c}{$0.894\pm0.026$} & 
\multicolumn{1}{|c}{$0.977\pm0.012$} & 
\multicolumn{1}{|c}{$0.824\pm0.009$} & 
\multicolumn{1}{|c|}{\bf{$0.990\pm0.002$}} \\
\hline 
\end{tabular}
\end{table}

\clearpage
\section{Figures}

\begin{figure}[ht]
\begin{center}
\includegraphics[width=14cm]{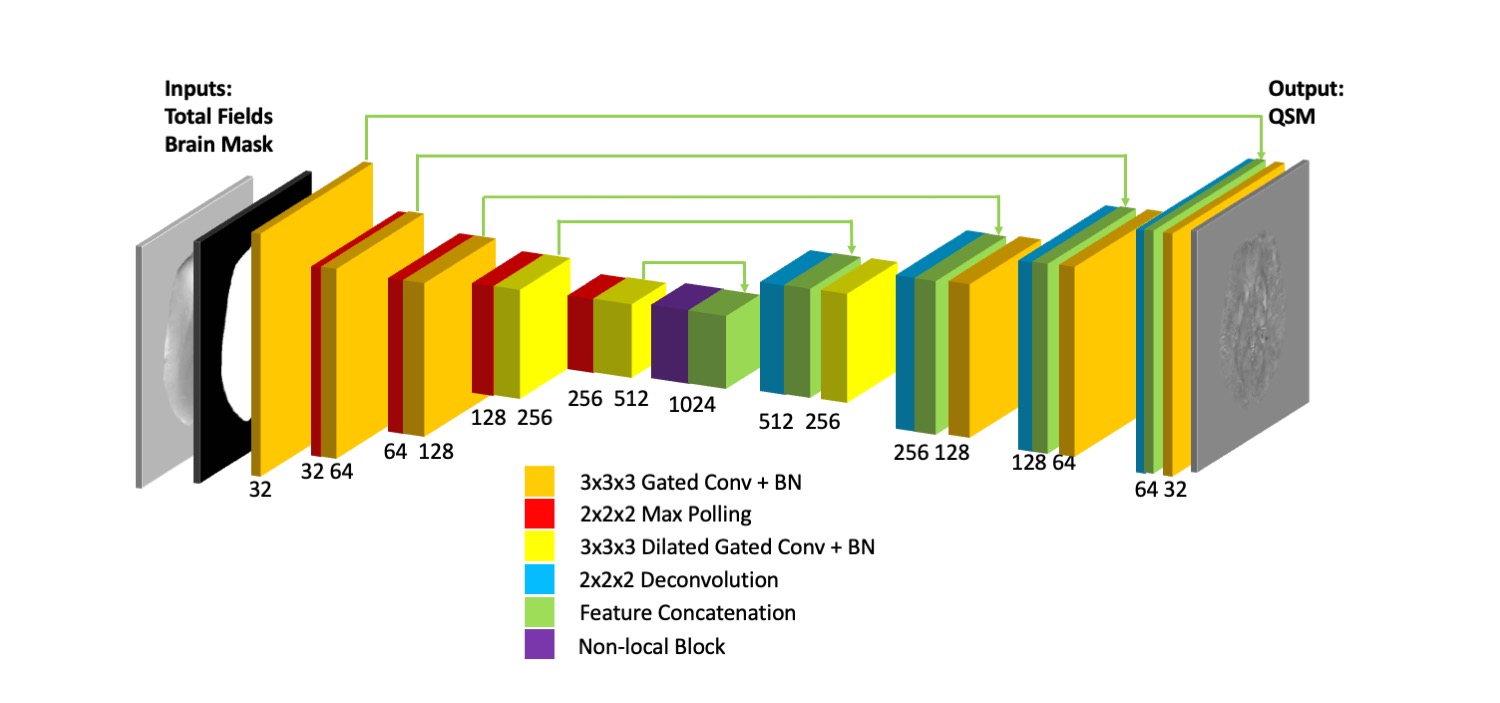}
\captionv{12}{}{Network structure of SSQSMnet. A 3D encoder-decoder architecture was designed with 6 gated convolutional layers (kernel size 3x3x3, dilated rate 1x1x1), 3 gated convolutional layers (kernel size 3x3x3, dilated rate 2x2x2), 4 max pooling layer (pool size 2x2x2, stride size 2x2x2), 1 non-local block, 4 deconvolution layers (kernel size 3x3x3, stride size 2x2x2), 9 normalization layers, 5 feature concatenations, and 1 convolutional layer (kernel size 3x3x3, linear activation). 
\label{fig_ssQSMNet}}
\end{center}
\end{figure}

\section*{References}
\addcontentsline{toc}{section}{\numberline{}References}
\vspace*{-20mm}
\bibliographystyle{unsrtnat}
\bibliography{reference.bib}

\begin{thebibliography}{46}
\providecommand{\natexlab}[1]{#1}
\providecommand{\url}[1]{\texttt{#1}}
\expandafter\ifx\csname urlstyle\endcsname\relax
  \providecommand{\doi}[1]{doi: #1}\else
  \providecommand{\doi}{doi: \begingroup \urlstyle{rm}\Url}\fi

\bibitem[Schenck(1996)]{schenck1996role}
John~F Schenck.
\newblock The role of magnetic susceptibility in magnetic resonance imaging:
  Mri magnetic compatibility of the first and second kinds.
\newblock \emph{Med. Phys.}, 23\penalty0 (6):\penalty0 815--850, 1996.

\bibitem[Haacke et~al.(2004)Haacke, Xu, Cheng, and
  Reichenbach]{haacke2004susceptibility}
E~Mark Haacke, Yingbiao Xu, Yu-Chung~N Cheng, and J{\"u}rgen~R Reichenbach.
\newblock Susceptibility weighted imaging (swi).
\newblock \emph{Magn. Res. Med.}, 52\penalty0 (3):\penalty0 612--618, 2004.

\bibitem[Wang and Liu(2015)]{wang2015quantitative}
Yi~Wang and Tian Liu.
\newblock Quantitative susceptibility mapping (qsm): decoding mri data for a
  tissue magnetic biomarker.
\newblock \emph{Magn. Res. Med.}, 73\penalty0 (1):\penalty0 82--101, 2015.

\bibitem[Babikian et~al.(2005)Babikian, Freier, Tong, Nickerson, Wall,
  Holshouser, Burley, Riggs, and Ashwal]{babikian2005susceptibility}
Talin Babikian, M~Catherin Freier, Karen~A Tong, Joshua~P Nickerson,
  Christopher~J Wall, Barbara~A Holshouser, Todd Burley, Matt~L Riggs, and
  Stephen Ashwal.
\newblock Susceptibility weighted imaging: neuropsychologic outcome and
  pediatric head injury.
\newblock \emph{Pediatr. Neurol.}, 33\penalty0 (3):\penalty0 184--194, 2005.

\bibitem[Huang et~al.(2015)Huang, Kuo, Tseng, Chen, Chiu, and
  Chen]{huang2015susceptibility}
Yen-Lin Huang, Ying-Sheng Kuo, Ying-Chi Tseng, David Yen-Ting Chen, Wen-Ta
  Chiu, and Chi-Jen Chen.
\newblock Susceptibility-weighted mri in mild traumatic brain injury.
\newblock \emph{Neurology}, 84\penalty0 (6):\penalty0 580--585, 2015.

\bibitem[Thomas et~al.(2008)Thomas, Somasundaram, Thamburaj, Kesavadas, Gupta,
  Bodhey, and Kapilamoorthy]{thomas2008clinical}
Bejoy Thomas, Sivaraman Somasundaram, Krishnamoorthy Thamburaj, Chandrasekharan
  Kesavadas, Arun~Kumar Gupta, Narendra~K Bodhey, and Tirur~Raman
  Kapilamoorthy.
\newblock Clinical applications of susceptibility weighted mr imaging of the
  brain--a pictorial review.
\newblock \emph{Neuroradiology}, 50\penalty0 (2):\penalty0 105--116, 2008.

\bibitem[De~Souza et~al.(2008)De~Souza, Domingues, Cruz, Domingues, Iasbeck,
  and Gasparetto]{de2008susceptibility}
JM~De~Souza, RC~Domingues, LCH Cruz, FS~Domingues, T~Iasbeck, and
  EL~Gasparetto.
\newblock Susceptibility-weighted imaging for the evaluation of patients with
  familial cerebral cavernous malformations: a comparison with t2-weighted fast
  spin-echo and gradient-echo sequences.
\newblock \emph{Am. J. Neuroradiol.}, 29\penalty0 (1):\penalty0 154--158, 2008.

\bibitem[Sehgal et~al.(2006)Sehgal, Delproposto, Haddar, Haacke, Sloan,
  Zamorano, Barger, Hu, Xu, Prabhakaran, et~al.]{sehgal2006susceptibility}
Vivek Sehgal, Zachary Delproposto, Djamel Haddar, E~Mark Haacke, Andrew~E
  Sloan, Lucia~J Zamorano, Geoffery Barger, Jiani Hu, Yingbiao Xu,
  Karthik~Praveen Prabhakaran, et~al.
\newblock Susceptibility-weighted imaging to visualize blood products and
  improve tumor contrast in the study of brain masses.
\newblock \emph{J. Mag. Reson. Im.}, 24\penalty0 (1):\penalty0 41--51, 2006.

\bibitem[Rauscher et~al.(2005)Rauscher, Sedlacik, Barth, Mentzel, and
  Reichenbach]{rauscher2005magnetic}
Alexander Rauscher, Jan Sedlacik, Markus Barth, Hans-Joachim Mentzel, and
  J{\"u}rgen~R Reichenbach.
\newblock Magnetic susceptibility-weighted mr phase imaging of the human brain.
\newblock \emph{Am. J. Neuroradiol.}, 26\penalty0 (4):\penalty0 736--742, 2005.

\bibitem[Haacke et~al.(2009)Haacke, Makki, Ge, Maheshwari, Sehgal, Hu, Selvan,
  Wu, Latif, Xuan, et~al.]{haacke2009characterizing}
E~Mark Haacke, Malek Makki, Yulin Ge, Megha Maheshwari, Vivek Sehgal, Jiani Hu,
  Madeswaran Selvan, Zhen Wu, Zahid Latif, Yang Xuan, et~al.
\newblock Characterizing iron deposition in multiple sclerosis lesions using
  susceptibility weighted imaging.
\newblock \emph{J. Mag. Reson. Im.}, 29\penalty0 (3):\penalty0 537--544, 2009.

\bibitem[Sehgal et~al.(2005)Sehgal, Delproposto, Haacke, Tong, Wycliffe, Kido,
  Xu, Neelavalli, Haddar, and Reichenbach]{sehgal2005clinical}
Vivek Sehgal, Zachary Delproposto, E~Mark Haacke, Karen~A Tong, Nathaniel
  Wycliffe, Daniel~K Kido, Yingbiao Xu, Jaladhar Neelavalli, Djamel Haddar, and
  J{\"u}rgen~R Reichenbach.
\newblock Clinical applications of neuroimaging with susceptibility-weighted
  imaging.
\newblock \emph{J. Mag. Reson. Im.}, 22\penalty0 (4):\penalty0 439--450, 2005.

\bibitem[Haacke et~al.(2005)Haacke, Cheng, House, Liu, Neelavalli, Ogg, Khan,
  Ayaz, Kirsch, and Obenaus]{haacke2005imaging}
E~Mark Haacke, Norman~YC Cheng, Michael~J House, Qiang Liu, Jaladhar
  Neelavalli, Robert~J Ogg, Asadullah Khan, Muhammad Ayaz, Wolff Kirsch, and
  Andre Obenaus.
\newblock Imaging iron stores in the brain using magnetic resonance imaging.
\newblock \emph{Magn. Reson. Imaging}, 23\penalty0 (1):\penalty0 1--25, 2005.

\bibitem[Deistung et~al.(2013{\natexlab{a}})Deistung, Sch{\"a}fer, Schweser,
  Biedermann, Turner, and Reichenbach]{deistung2013toward}
Andreas Deistung, Andreas Sch{\"a}fer, Ferdinand Schweser, Uta Biedermann,
  Robert Turner, and J{\"u}rgen~R Reichenbach.
\newblock Toward in vivo histology: a comparison of quantitative susceptibility
  mapping (qsm) with magnitude-, phase-, and r2⁎-imaging at ultra-high
  magnetic field strength.
\newblock \emph{NeuroImage}, 65:\penalty0 299--314, 2013{\natexlab{a}}.

\bibitem[Zhang et~al.(2015)Zhang, Liu, Gupta, Spincemaille, Nguyen, and
  Wang]{zhang2015quantitative}
Jingwei Zhang, Tian Liu, Ajay Gupta, Pascal Spincemaille, Thanh~D Nguyen, and
  Yi~Wang.
\newblock Quantitative mapping of cerebral metabolic rate of oxygen (cmro2)
  using quantitative susceptibility mapping (qsm).
\newblock \emph{Magn. Res. Med.}, 74\penalty0 (4):\penalty0 945--952, 2015.

\bibitem[Zheng et~al.(2013)Zheng, Nichol, Liu, Cheng, and
  Haacke]{zheng2013measuring}
Weili Zheng, Helen Nichol, Saifeng Liu, Yu-Chung~N Cheng, and E~Mark Haacke.
\newblock Measuring iron in the brain using quantitative susceptibility mapping
  and x-ray fluorescence imaging.
\newblock \emph{NeuroImage}, 78:\penalty0 68--74, 2013.

\bibitem[Deistung et~al.(2013{\natexlab{b}})Deistung, Schweser, Wiestler,
  Abello, Roethke, Sahm, Wick, Nagel, Heiland, Schlemmer,
  et~al.]{deistung2013quantitative}
Andreas Deistung, Ferdinand Schweser, Benedikt Wiestler, Mario Abello, Matthias
  Roethke, Felix Sahm, Wolfgang Wick, Armin~Michael Nagel, Sabine Heiland,
  Heinz-Peter Schlemmer, et~al.
\newblock Quantitative susceptibility mapping differentiates between blood
  depositions and calcifications in patients with glioblastoma.
\newblock \emph{PloS One}, 8\penalty0 (3):\penalty0 e57924, 2013{\natexlab{b}}.

\bibitem[Wisnieff et~al.(2015)Wisnieff, Ramanan, Olesik, Gauthier, Wang, and
  Pitt]{wisnieff2015quantitative}
Cynthia Wisnieff, Sriram Ramanan, John Olesik, Susan Gauthier, Yi~Wang, and
  David Pitt.
\newblock Quantitative susceptibility mapping (qsm) of white matter multiple
  sclerosis lesions: interpreting positive susceptibility and the presence of
  iron.
\newblock \emph{Magn. Res. Med.}, 74\penalty0 (2):\penalty0 564--570, 2015.

\bibitem[Bilgic et~al.(2012)Bilgic, Pfefferbaum, Rohlfing, Sullivan, and
  Adalsteinsson]{bilgic2012mri}
Berkin Bilgic, Adolf Pfefferbaum, Torsten Rohlfing, Edith~V Sullivan, and Elfar
  Adalsteinsson.
\newblock Mri estimates of brain iron concentration in normal aging using
  quantitative susceptibility mapping.
\newblock \emph{NeuroImage}, 59\penalty0 (3):\penalty0 2625--2635, 2012.

\bibitem[Lotfipour et~al.(2012)Lotfipour, Wharton, Schwarz, Gontu, Sch{\"a}fer,
  Peters, Bowtell, Auer, Gowland, and Bajaj]{lotfipour2012high}
Ashley~K Lotfipour, Samuel Wharton, Stefan~T Schwarz, V~Gontu, Andreas
  Sch{\"a}fer, Andrew~M Peters, Richard~W Bowtell, Dorothee~P Auer, Penny~A
  Gowland, and Nin~PS Bajaj.
\newblock High resolution magnetic susceptibility mapping of the substantia
  nigra in parkinson's disease.
\newblock \emph{J. Magn. Reson Im.}, 35\penalty0 (1):\penalty0 48--55, 2012.

\bibitem[Sharma et~al.(2015)Sharma, Hernando, Horng, and
  Reeder]{sharma2015quantitative}
Samir~D Sharma, Diego Hernando, Debra~E Horng, and Scott~B Reeder.
\newblock Quantitative susceptibility mapping in the abdomen as an imaging
  biomarker of hepatic iron overload.
\newblock \emph{Magn. Res. Med.}, 74\penalty0 (3):\penalty0 673--683, 2015.

\bibitem[Fan et~al.(2014)Fan, Bilgic, Gagnon, Witzel, Bhat, Rosen, and
  Adalsteinsson]{fan2014quantitative}
Audrey~P Fan, Berkin Bilgic, Louis Gagnon, Thomas Witzel, Himanshu Bhat,
  Bruce~R Rosen, and Elfar Adalsteinsson.
\newblock Quantitative oxygenation venography from mri phase.
\newblock \emph{Magn. Res. Med.}, 72\penalty0 (1):\penalty0 149--159, 2014.

\bibitem[Koch et~al.(2018)Koch, Meier, Karr, Nencka, Muftuler, and
  McCrea]{koch2018quantitative}
Kevin~M Koch, Timothy~B Meier, Robin Karr, Andrew~S Nencka, L~Tugan Muftuler,
  and Michael McCrea.
\newblock Quantitative susceptibility mapping after sports-related concussion.
\newblock \emph{Am. J. Neuroradiol.}, 39\penalty0 (7):\penalty0 1215--1221,
  2018.

\bibitem[Schweser et~al.(2011)Schweser, Deistung, Lehr, and
  Reichenbach]{schweser2011quantitative}
Ferdinand Schweser, Andreas Deistung, Berengar~Wendel Lehr, and
  J{\"u}rgen~Rainer Reichenbach.
\newblock Quantitative imaging of intrinsic magnetic tissue properties using
  mri signal phase: an approach to in vivo brain iron metabolism?
\newblock \emph{NeuroImage}, 54\penalty0 (4):\penalty0 2789--2807, 2011.

\bibitem[Sun and Wilman(2014)]{sun2014background}
Hongfu Sun and Alan~H Wilman.
\newblock Background field removal using spherical mean value filtering and
  tikhonov regularization.
\newblock \emph{Magn. Res. Med.}, 71\penalty0 (3):\penalty0 1151--1157, 2014.

\bibitem[Liu et~al.(2011)Liu, Khalidov, de~Rochefort, Spincemaille, Liu,
  Tsiouris, and Wang]{liu2011novel}
Tian Liu, Ildar Khalidov, Ludovic de~Rochefort, Pascal Spincemaille, Jing Liu,
  A~John Tsiouris, and Yi~Wang.
\newblock A novel background field removal method for mri using projection onto
  dipole fields.
\newblock \emph{NMR Biomed.}, 24\penalty0 (9):\penalty0 1129--1136, 2011.

\bibitem[Zhou et~al.(2014)Zhou, Liu, Spincemaille, and
  Wang]{zhou2014background}
Dong Zhou, Tian Liu, Pascal Spincemaille, and Yi~Wang.
\newblock Background field removal by solving the laplacian boundary value
  problem.
\newblock \emph{NMR Biomed.}, 27\penalty0 (3):\penalty0 312--319, 2014.

\bibitem[Berkin et~al.(2014)Berkin, Christian, Lawrence, and
  Kawin]{Berkin2014SingleStep}
Bilgic Berkin, Langkammer Christian, L~Wald Lawrence, and Setsompop Kawin.
\newblock Single-step qsm with fast reconstruction.
\newblock \emph{ISMRM}, 3321, 2014.

\bibitem[Kristian et~al.(2014)Kristian, Stefan, Benedikt, Markus, and
  Christian]{Kristian2014SingleStep}
Bredies Kristian, Ropele Stefan, A~Poser Benedikt, Barth Markus, and Langkammer
  Christian.
\newblock Single-step quantitative susceptibility mapping using total
  generalized variation and 3d epi.
\newblock \emph{ISMRM}, 0604, 2014.

\bibitem[Liu et~al.(2014)Liu, Zhou, Spincemaille, and
  Wang]{Tian2014Differential}
Tian Liu, Dong Zhou, Pascal Spincemaille, and Yi~Wang.
\newblock Differential approach to quantitative susceptibility mapping without
  background field removal.
\newblock \emph{ISMRM}, 0597, 2014.

\bibitem[Sharma et~al.(2014)Sharma, Hernando, Horng, and
  Reeder]{Samir2014SingleStep}
Samir~D Sharma, Diego Hernando, Debra~E Horng, and Scott~B Reeder.
\newblock A joint background field removal and dipole deconvolution approach
  for quantitative susceptibility mapping in the liver.
\newblock \emph{ISMRM}, 0606, 2014.

\bibitem[Liu et~al.(2017)Liu, Kee, Zhou, Wang, and
  Spincemaille]{liu2017preconditioned}
Zhe Liu, Youngwook Kee, Dong Zhou, Yi~Wang, and Pascal Spincemaille.
\newblock Preconditioned total field inversion (tfi) method for quantitative
  susceptibility mapping.
\newblock \emph{Magn. Res. Med.}, 78\penalty0 (1):\penalty0 303--315, 2017.

\bibitem[Chatnuntawech et~al.(2017)Chatnuntawech, McDaniel, Cauley, Gagoski,
  Langkammer, Martin, Grant, Wald, Setsompop, Adalsteinsson,
  et~al.]{chatnuntawech2017single}
Itthi Chatnuntawech, Patrick McDaniel, Stephen~F Cauley, Borjan~A Gagoski,
  Christian Langkammer, Adrian Martin, P~Ellen Grant, Lawrence~L Wald, Kawin
  Setsompop, Elfar Adalsteinsson, et~al.
\newblock Single-step quantitative susceptibility mapping with variational
  penalties.
\newblock \emph{NMR Biomed.}, 30\penalty0 (4):\penalty0 e3570, 2017.

\bibitem[Sun et~al.(2018)Sun, Ma, MacDonald, and Pike]{sun2018whole}
Hongfu Sun, Yuhan Ma, M~Ethan MacDonald, and G~Bruce Pike.
\newblock Whole head quantitative susceptibility mapping using a least-norm
  direct dipole inversion method.
\newblock \emph{NeuroImage}, 2018.

\bibitem[Marques and Bowtell(2005)]{marques2005application}
JP~Marques and R~Bowtell.
\newblock Application of a fourier-based method for rapid calculation of field
  inhomogeneity due to spatial variation of magnetic susceptibility.
\newblock \emph{Concepts Magn. Reson. B.}, 25\penalty0 (1):\penalty0 65--78,
  2005.

\bibitem[Salomir et~al.(2003)Salomir, de~Senneville, and
  Moonen]{salomir2003fast}
Rares Salomir, Baudouin~Denis de~Senneville, and Chrit~TW Moonen.
\newblock A fast calculation method for magnetic field inhomogeneity due to an
  arbitrary distribution of bulk susceptibility.
\newblock \emph{Concepts Magn. Reson. B.}, 19\penalty0 (1):\penalty0 26--34,
  2003.

\bibitem[Liu et~al.(2009)Liu, Spincemaille, De~Rochefort, Kressler, and
  Wang]{liu2009calculation}
Tian Liu, Pascal Spincemaille, Ludovic De~Rochefort, Bryan Kressler, and
  Yi~Wang.
\newblock Calculation of susceptibility through multiple orientation sampling
  (cosmos): a method for conditioning the inverse problem from measured
  magnetic field map to susceptibility source image in mri.
\newblock \emph{Magn. Res. Med.}, 61\penalty0 (1):\penalty0 196--204, 2009.

\bibitem[Liu(2010)]{liu2010susceptibility}
Chunlei Liu.
\newblock Susceptibility tensor imaging.
\newblock \emph{Magn. Res. Med.}, 63\penalty0 (6):\penalty0 1471--1477, 2010.

\bibitem[Ashburner and Friston(2005)]{ashburner2005unified}
John Ashburner and Karl~J Friston.
\newblock Unified segmentation.
\newblock \emph{NeuroImage}, 26\penalty0 (3):\penalty0 839--851, 2005.

\bibitem[Liu et~al.(2019)Liu, Nencka, Muftuler, Swearingen, Karr, and
  Koch]{liu2019quantitative}
Juan Liu, Andrew~S Nencka, L~Tugan Muftuler, Brad Swearingen, Robin Karr, and
  Kevin~M Koch.
\newblock Quantitative susceptibility inversion through parcellated
  multiresolution neural networks and k-space substitution.
\newblock \emph{arXiv}, 2019.

\bibitem[Wang et~al.(2018)Wang, Girshick, Gupta, and He]{wang2018non}
Xiaolong Wang, Ross Girshick, Abhinav Gupta, and Kaiming He.
\newblock Non-local neural networks.
\newblock In \emph{IEEE/CVF}, pages 7794--7803, 2018.

\bibitem[Liu et~al.(2012)Liu, Liu, de~Rochefort, Ledoux, Khalidov, Chen,
  Tsiouris, Wisnieff, Spincemaille, Prince, et~al.]{liu2012morphology}
Jing Liu, Tian Liu, Ludovic de~Rochefort, James Ledoux, Ildar Khalidov, Weiwei
  Chen, A~John Tsiouris, Cynthia Wisnieff, Pascal Spincemaille, Martin~R
  Prince, et~al.
\newblock Morphology enabled dipole inversion for quantitative susceptibility
  mapping using structural consistency between the magnitude image and the
  susceptibility map.
\newblock \emph{NeuroImage}, 59\penalty0 (3):\penalty0 2560--2568, 2012.

\bibitem[Bilgic et~al.(2015)Bilgic, Gagoski, Cauley, Fan, Polimeni, Grant,
  Wald, and Setsompop]{bilgic2015wave}
Berkin Bilgic, Borjan~A Gagoski, Stephen~F Cauley, Audrey~P Fan, Jonathan~R
  Polimeni, P~Ellen Grant, Lawrence~L Wald, and Kawin Setsompop.
\newblock Wave-caipi for highly accelerated 3d imaging.
\newblock \emph{Magn. Res. Med.}, 73\penalty0 (6):\penalty0 2152--2162, 2015.

\bibitem[Langkammer et~al.(2018)Langkammer, Schweser, Shmueli, Kames, Li, Guo,
  Milovic, Kim, Wei, Bredies, et~al.]{langkammer2018quantitative}
Christian Langkammer, Ferdinand Schweser, Karin Shmueli, Christian Kames,
  Xu~Li, Li~Guo, Carlos Milovic, Jinsuh Kim, Hongjiang Wei, Kristian Bredies,
  et~al.
\newblock Quantitative susceptibility mapping: report from the 2016
  reconstruction challenge.
\newblock \emph{Magn. Res. Med.}, 79\penalty0 (3):\penalty0 1661--1673, 2018.

\bibitem[Yi(2019)]{MEDIcornellWEB}
Wang Yi.
\newblock {Cornell MRI Research Lab}, 2019.
\newblock URL \url{http://weill.cornell.edu/mri/pages/qsm.html}.

\bibitem[Berkin(2019)]{berkinMITWEB}
Bilgic Berkin.
\newblock {Berkin Bilgic Personal Website}, 2019.
\newblock URL \url{http://martinos.org/~berkin/software.html}.

\bibitem[Hongfu(2019)]{LNQSMCode}
Sun Hongfu.
\newblock {LN-QSM Github}, 2019.
\newblock URL \url{https://github.com/sunhongfu/QSM}.

\end{thebibliography}

\end{document}